\documentclass[sn-mathphys-num]{sn-jnl}% Math and Physical Sciences Numbered Reference Style 
\usepackage{graphicx}%
\usepackage{multirow}%
\usepackage{amsmath,amssymb,amsfonts}%
\usepackage{amsthm}%
\usepackage[title]{appendix}%
\usepackage{xcolor}%
\usepackage{textcomp}%
\usepackage{manyfoot}%
\usepackage{booktabs}%
\usepackage{lmodern}
\usepackage{listings}%
\usepackage{booktabs}
\begin{document}

\title[Article Title]{Selection of Supervised Learning-based Sparse Matrix Reordering Algorithms}


\author*[1]{\fnm{Tao} \sur{Tang}}\email{taotang84@nudt.edu.cn}

\author[1]{\fnm{Youfu} \sur{Jiang}}

\author[1]{\fnm{Yingbo} \sur{Cui}}

\author[1]{\fnm{Jianbin} \sur{Fang}}
\author[1]{\fnm{Peng} \sur{Zhang}}
\author[1]{\fnm{Lin} \sur{Peng}}
\author[1]{\fnm{Chun} \sur{Huang}}

\affil*[1]{\orgdiv{College of Computer Science and Technology}, \orgname{National University of Defense Technology}, \city{Changsha}, \postcode{410073}, \state{Hunan}, \country{China}}


\abstract{Sparse matrix ordering is a vital optimization technique often employed for solving large-scale sparse matrices. Its goal is to minimize the matrix bandwidth by reorganizing its rows and columns, thus enhancing efficiency. Conventional methods for algorithm selection usually depend on brute-force search or empirical knowledge, lacking the ability to adjust to diverse sparse matrix structures. As a result, we have introduced a supervised learning-based model for choosing sparse matrix reordering algorithms. This model grasps the correlation between matrix characteristics and commonly utilized reordering algorithms, facilitating the automated and intelligent selection of the suitable sparse matrix reordering algorithm. Experiments conducted on the Florida sparse matrix dataset reveal that our model can accurately predict the optimal reordering algorithm for various matrices, leading to a 55.37\% reduction in solution time compared to solely using the AMD reordering algorithm, with an average speedup ratio of 1.45.}




\maketitle

\section{Introduction}\label{sec1}

In various numerical simulation fields such as fluid dynamics, structural mechanics, and quantum mechanics, many practical application problems ultimately boil down to solving linear equation systems in the form of Ax = B, where most coefficient matrices are sparse matrices. Currently, there are two main approaches to solving these sparse linear equation systems: direct methods and iterative methods. Direct methods typically involve operating directly on the matrix and vector elements to solve the linear equation system. These methods often exploit the special structure and properties of the matrix, such as triangular matrices or banded matrices, to efficiently solve sparse linear equation systems. They are suitable for medium-sized sparse matrix linear equation systems, offering high accuracy and stability, with popular solvers including MUMPS, SuperLU, and HSL. On the other hand, iterative methods start with an initial solution vector and alliterative refine it to approximate the true solution of the sparse matrix linear equation system. These methods are suitable for large-scale sparse matrix linear equation systems where high solution accuracy is not required, offering low memory requirements and computational complexity, with popular solvers including PETSc and IML.

For sparse direct solving, the most crucial step involves factorizing the sparse matrix, which is also one of the most time-consuming and memory-intensive stages in solving sparse matrices. Particularly for large-scale sparse matrices, factorization demands significant computational resources and memory usage, significantly impacting computational efficiency and scalability. To decrease the memory footprint and computational time required for matrix factorization, a common optimization approach is to reorder the matrix\cite{trotter2023bringing} Through reordering the sparse matrix, the computation can better utilize cache, parallel computing, and vectored instructions, effectively reducing the computational overhead of direct solvers. Additionally, reordering aids in decreasing the number of fill-in elements generated during factorization, thereby reducing storage requirements and enhancing algorithmic efficiency.

However, finding a reordering algorithm that minimizes fill-in is an NP-hard problem\cite{dasgupta2023alpha}. Many researchers have investigated reordering strategies, and Tinney \textit{et al.} \cite{tinney1967direct} were among the first to apply the minimum-degree (MD) algorithm to sparse matrix solving to improve solution efficiency, which is one of the most widely used heuristic reordering algorithms. The core idea is to represent the sparse matrix as an adjacency graph, select a node with the minimum degree, and then eliminate nodes in a certain order until all nodes in the matrix are processed, resulting in the final reordering result. Later, Amestoy \textit{et al.} \cite{amestoy1996approximate} built upon the minimum-degree algorithm and proposed the approximate minimum-degree (AMD) algorithm, significantly improving the time and space complexity while maintaining similar reordering quality. The Cuthill-McKee (CM) algorithm \cite{cuthill1969reducing}, first introduced by Cuthill and McKee, uses breadth-first search to traverse the matrix and renumber the nodes, resulting in a reordered matrix with smaller bandwidth. Sherman \textit{et al.} \cite{liu1976comparative} reversed the CM algorithm's reordering result to obtain the more commonly used Reverse Cuthill-McKee (RCM) algorithm, which has been shown to perform better than CM on linear equation systems. In addition, another popular sparse matrix reordering algorithm is the nested dissection (ND) algorithm \cite{george1973nested}, first proposed by Alan George, which uses a divide-and-conquer strategy and has been proven to have good reordering effects on certain types of graphs \cite{khaira1992nested}.

There are currently many sparse matrix reordering algorithms available, but each has its applicable range and no single algorithm can adapt to all types of sparse matrices. It may even slow down the matrix solution process if used improperly. Therefore, selecting the proper sparse matrix reordering algorithm is a worthy research topic. Traditional methods usually rely on experience or brute-force search to select a suitable algorithm, often facing issues such as low efficiency and insufficient accuracy.

To address this problem, we proposed a supervised learning-based sparse matrix reordering algorithm selection model. For a given sparse matrix, the model extracts sparse matrix features and uses machine learning models to predict the effectiveness of various reordering algorithms, As a result, automated and intelligent algorithm selection is achieved, enhancing the efficiency of solving sparse matrices.

The main contributions of this paper are as follows:

(1) We conducted experiments on over 900 sparse matrices to test multiple reordering algorithms and analyzed their impact on sparse matrix solution performance. The results show that some sparse matrices exhibit significant differences in solution times under different reordering algorithms, and no single algorithm can adapt to all sparse matrix structures.

(2) We proposed a machine learning model for automatically selecting sparse matrix reordering algorithms. First, we selected the optimal label for each matrix based on the shortest solution time and extracted predefined features from each matrix to form a dataset. Then, we divided the dataset into training and testing sets, trained and tested multiple machine learning algorithms, and evaluated the model using accuracy as the evaluation metric. We optimized the hyperparameters with grid search and k-fold cross-validation and selected the machine learning algorithm with the highest accuracy to predict the optimal reordering algorithm for sparse matrix solution.

(3) The experiments on the testing set show that our model achieves a high accuracy rate of 86.7\%. Compared to the default AMD reordering algorithm used by many direct solvers, our method significantly improves solution performance. Specifically, the predicted reordering algorithm reduces solution times by 55.37\% compared to the AMD algorithm alone, and the average speedup ratio is 1.45 for all matrices.
\section{Analysis of Reordering Algorithm Performance
}\label{sec10}
Different reordering algorithms have different optimization goals, so using different reordering algorithms for sparse matrix solving may result in significant differences. We selected several matrices with over 100,000 non-zero elements and uses four reordering algorithms (AMD, SCOTCH, ND, RCM) in the MUMPS solver to solve them, with the results shown in Table 1. We can observe that for the same matrix,, the solution times differ greatly under different reordering algorithms, with differences of up to several thousand times, such as the lhr07c matrix. Therefore, the performance of matrix solving heavily on the selection of ordering algorithm.

\begin{table}[!t]
\caption{Matrix Solution Times with Various Reordering Algorithms}\label{tab1}%
\begin{tabular}{@{}ccccccc@{}}
\toprule
Matrix Name & AMD(s)     & SCOTCH(s)  & ND(s)       & RCM(s)      & Nnz     & Dimension \\ \midrule
ASIC\_320k  & 141.708 & 21.1478 & 1.2294   & 3.0100   & 2635364 & 321821    \\
pf2177      & 18.0631 & 13.1600 & 172.8088 & 174.634  & 725144  & 9728      \\
crystk02    & 4.3335  & 2.4262  & 20.7101  & 15.1643  & 968583  & 13965     \\
SiH4        & 8.7350  & 3.6215  & 28.8439  & 28.0424  & 171903  & 5041      \\
obstclae    & 0.3100  & 0.6710  & 13.7744  & 77.7252  & 197608  & 40000     \\
lhr07c      & 0.1449  & 0.4079  & 154.999  & 258.299  & 156508  & 7337      \\
nemeth17    & 0.2886  & 0.3519  & 0.1816   & 0.0902   & 629620  & 9506      \\
af23560     & 1.9415  & 2.9312  & 29.4057  & 328.153  & 484256  & 23560     \\
pli         & 58.4708 & 17.4346 & 198.9748 & 215.6674 & 1350309 & 22695     \\ \bottomrule
\end{tabular}
\end{table}

To more intuitively illustrate the relationship between sparse matrix solution times and reordering algorithms, we randomly selected 30 matrices from the sparse matrix collection, normalized the solution times for each matrix under different algorithms, and used a heatmap to display the results. The darker the color, the shorter the solution time, and vice versa. The specific results are shown in Fig. 1. From the figure, we can see that the same reordering algorithm performs differently on different matrices, with varying shades of color, indicating that the suitability of a reordering algorithm is not fixed. We can also observe that for most matrices, the AMD reordering algorithm generally performs well, indirectly reflecting the widespread applicability and effectiveness of the AMD algorithm.
\begin{figure}[htbp]
\centering
\includegraphics[width=0.9\textwidth]{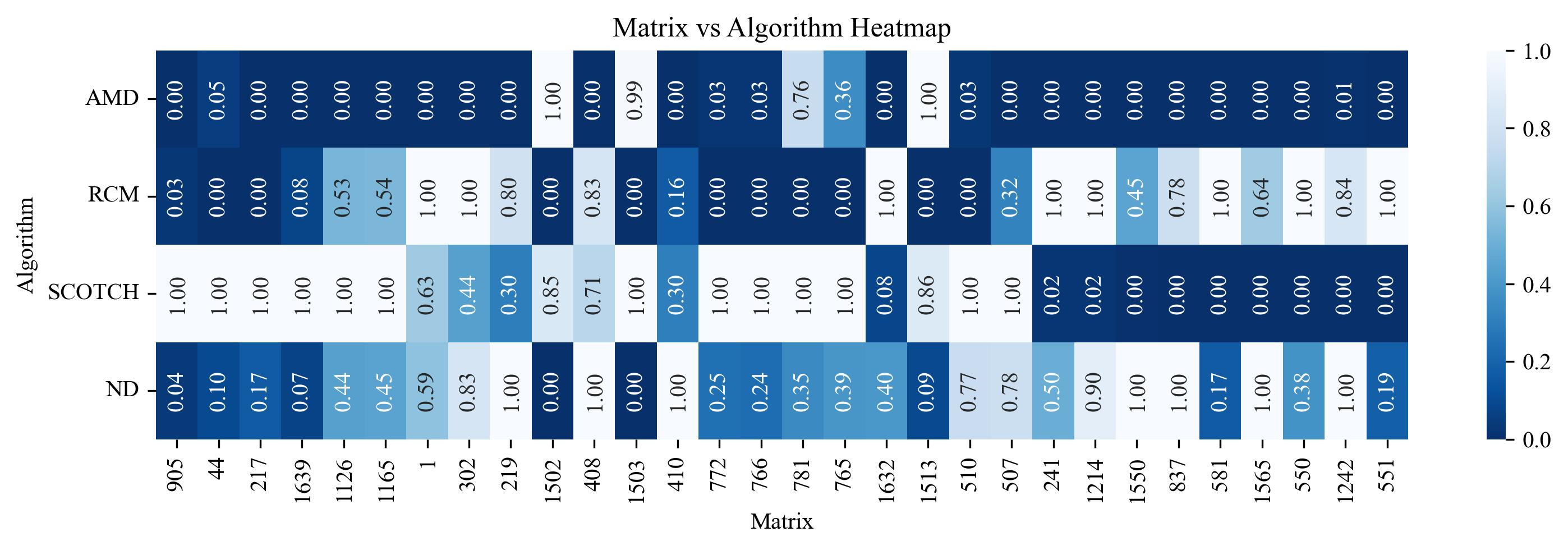}
%\captionsetup\caption
\caption{Comparison of Solution Times for Sparse Matrix Reordering Algorithms}\label{fig1}
\end{figure}

In conclusion, blindly applying a single sparse matrix reordering algorithm is not recommended for solving sparse matrices. Instead, it is often necessary to selectively choose a reordering algorithm based on the structure of the sparse matrix to enhance the efficiency of solving sparse matrices.

\section{Establishment of the Model
}\label{sec3}

\subsection{Methodology
}\label{subsec1}

The process flowchart of the sparse matrix reordering algorithm prediction model built in this paper is shown in Fig. 2. First, Python scripts are used to extract matrix features and statistics for different reordering algorithms, and the solution times for each matrix using different reordering algorithms are recorded. The optimal reordering algorithm label is marked for each matrix, and the pre-defined matrix features and optimal reordering algorithm labels are input into different machine learning models for training. Grid search is used to optimize related hyperparameters, and finally, the machine learning model with the highest prediction accuracy is trained. When making predictions, only the features of the matrix to be predicted need to be extracted and input into the trained model to obtain the predicted algorithm result.
\begin{figure}[htbp]
\centering
\includegraphics[width=0.9\textwidth]{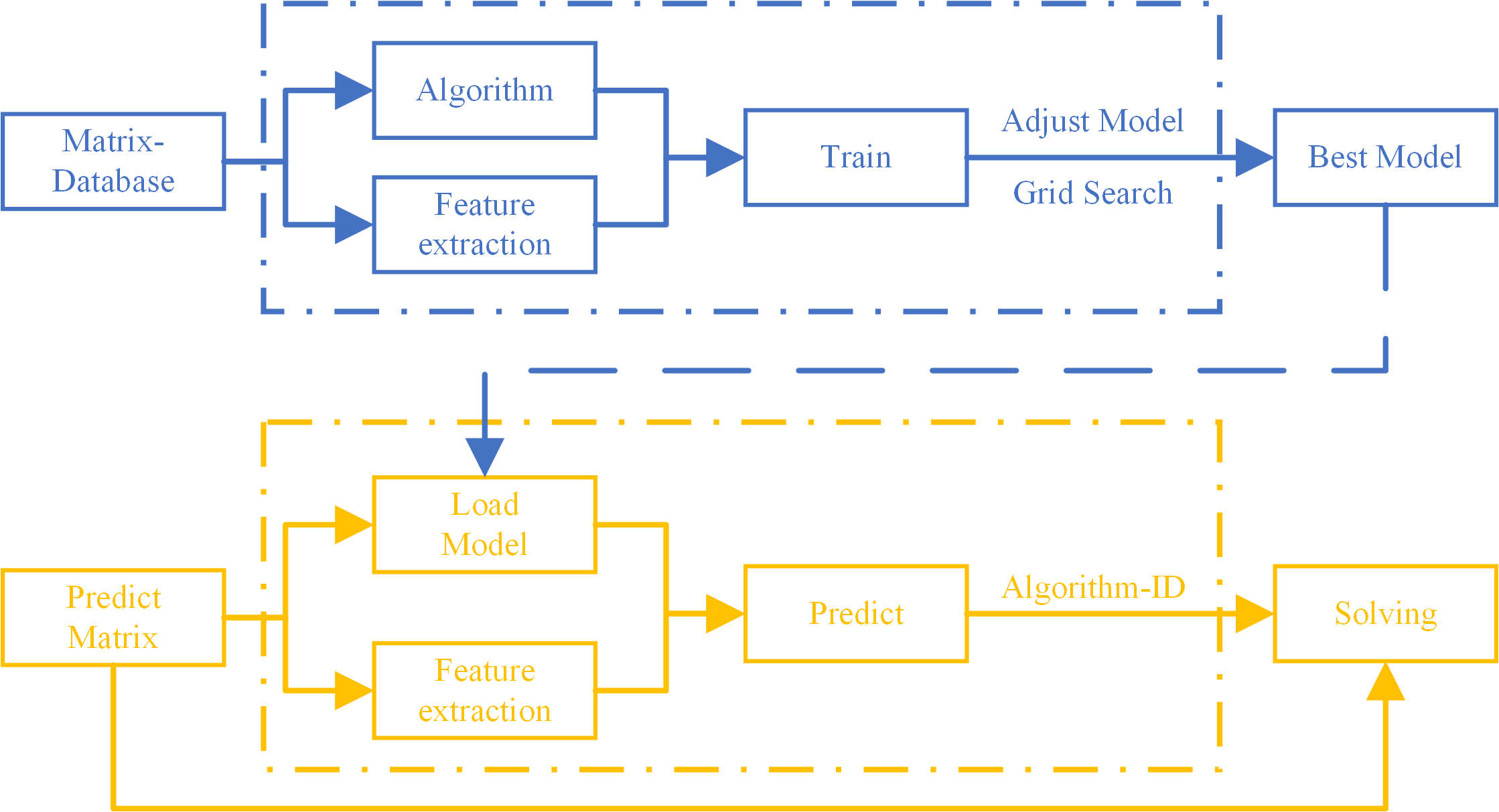}
%\captionsetup\caption
\caption{Sparse Matrix Reordering Algorithm Prediction Model}\label{fig2}
\end{figure}

In essence, our machine learning model aims to train a system that can predict the optimal reordering algorithm for a sparse input matrix, establishing a mapping between the matrix features and the best algorithm. This mapping is represented by the following formula:
\begin{equation}
f\left(\vec{x}_1, \vec{x}_2, \ldots, \vec{x}_n\right) \to y_n\text{ (AMD, SCOTCH, ND, RCM)}
\end{equation}

Where $\overrightarrow{x}_{i}$  represents the feature vector of the 
 $\text{i}$th sparse matrix in the training set, $y_{n}$ represents the target vector, which contains the indices of 4 sparse matrix reordering algorithms.

\subsection{Data Preprocessing}\label{subsec7}
This article selects the first 2000 matrices from the Florida Matrix Collection \cite{davis2011university}, which covers multiple fields such as fluid mechanics, electromagnetics, and quantum chemistry, and is a widely recognized sparse matrix collection in the public domain, widely used in developing and evaluating sparse matrix-related algorithms.

To obtain a suitable dataset for model training, we performed several preprocessing steps. First, the downloaded matrices were filtered to select only square matrices, and complex-type matrices were removed, leaving only real-type matrices. Then, Python scripts were used to randomly generate right-hand side vectors and solution vectors for each matrix, preparing them for subsequent sparse matrix solving.

The sparse matrix reordering algorithm has a significant impact on the performance of sparse matrix solving. MUMPS is a software package for solving linear systems of equations in the form of AX = B, suitable for large-scale solving problems and parallel computing environments. It provides various sparse matrix reordering algorithms, including AMD, AMF, SCOTCH, PORD, and QAMD, and supports custom input reordering results for sparse matrix solving. Therefore, when building the model training dataset, this article chooses MUMPS as the sparse matrix solver and uses the shortest solving time as the label for each matrix during MUMPS solving. In addition to the above-mentioned sparse matrix reordering algorithms, this article also includes RCM and ND reordering algorithms, which are implemented using the Scipy library and Metis graph partitioning software package \cite{karypis1998fast}, respectively. The reordering results of these two algorithms will be specified as input when using MUMPS to solve the sparse matrices. Then, the solving times of each matrix under different reordering algorithms are recorded, resulting in a total of 936 matrices' solving times. To better classify the reordering algorithms, inspired by the classification method of Trotter et al. \cite{trotter2023bringing}, we divided the seven reordering algorithms into four categories based on their optimization goals: bandwidth reduction algorithms, fill-in reduction algorithms, graph-based algorithms, and hybrid algorithms combining fill-in reduction and graph-based methods. Table 2 shows these algorithms.

% \usepackage{booktabs}
\begin{table}[]
\caption{Classification of Reordering Algorithms}\label{tab2}%
\begin{tabular}{@{}cc@{}}
\toprule
Category Names                                                                                                   & Reordering Algorithm \\ \midrule
bandwidth reduction algorithms                                                                                   & RCM                  \\
fill-in reduction algorithms                                                                                     & AMD,AMF,QAMD         \\
graph-based algorithms                                                                                           & ND                   \\
\begin{tabular}[c]{@{}c@{}}hybrid algorithms combining \\ fill-in reduction and graph-based methods\end{tabular} & SCOTCH,PORD          \\ \bottomrule
\end{tabular}
\end{table}

In the data preprocessing process, besides recording the solving time of the matrices under different reordering algorithms, we also need to take the reordering algorithm with the shortest solving time for each matrix as its label. Based on the classification method in Table 1, we finally selected the reordering algorithms with better solving performance and wider usage in each category, and ultimately used RCM, AMD, ND, and SCOTCH as the predictive labels for the matrices.
 \subsection{ Feature Data Selection and Construction
}\label{subsec3}
Due to the different optimization goals of various sparse matrix reordering algorithms, we selected the matrix features that are heavily dependent on sparse matrix reordering. We chose the bandwidth and profile of the matrix as the input features of the matrix. For a sparse matrix of dimension is $\text{N*N}$, the bandwidth refers to the maximum distance from the non-zero elements in each row to the main diagonal line, which can be expressed by the following formula:
\begin{equation}\begin{array}{rl}\text{Bandwidth}&=Max_{a_{i,j}\neq0}\big|i-j\big|\end{array}\end{equation}
The profile refers to the sum of the distances from the leftmost non-zero element in each row to the diagonal line, which can be expressed by the following formula:
\begin{equation}\overset{}{\text{Profile=}\sum_{i=1}^{N}i}-\min\{j|a_{i,j}\neq0\}\end{equation}
At the same time, during the statistical analysis of matrix solving time, it was discovered that the size of the matrix and the number of non-zero elements significantly impact matrix solving. As a result, the number of non-zero elements, matrix size, and other characteristics were chosen. Additionally, considering the optimization goals of some reordering algorithms, the node degree of the matrix was also chosen as an input feature, which refers to the number of non-zero elements connected to a node in the matrix. To effectively characterize the input sparse matrix, based on the above analysis, we selected 12 features, including the features mentioned above and their derived features, as inputs to the machine learning model. These features are shown in Table 3:

% \usepackage{booktabs}
\begin{table}[]
\caption{Matrix Features}\label{tab3}%
\begin{tabular}{@{}cc@{}}
\toprule
Feature     & Description                   \\ \midrule
dimension   & Number of Dimension           \\
nnz         & Number of Nonzero             \\
nnz\_ratio  & Proportion of Nonzero         \\
nnz\_max    & Max Number of Nonzero per Row \\
nnz\_min    & Min Number of Nonzero per Row \\
nnz\_avg    & Avg Number of Nonzero per Row \\
nnz\_std    & Std Number of Nonzero per Row \\
degree\_max & Max Node Degree               \\
degree\_min & Min Node Degree               \\
degree\_avg & Avg Node Degree               \\
bandwidth   & Bandwidth of Sparse Matrix    \\
profile     & Profile of Sparse Matrix      \\ \bottomrule
\end{tabular}
\end{table}
Before model training, we process the sparse matrix data and obtain the corresponding features for each sparse matrix through self-written Python code.
 \subsection{Model Building and Training
}\label{subsec4}
Scikit-learn is a Python library for machine learning and data mining, providing a rich set of tools and functions for various common machine learning tasks, including classification, regression, clustering, dimensionality reduction, and more. Therefore, in this article, we used Scikit-learn to build and train models, selecting seven algorithm models provided by Scikit-learn, including Random Forest Algorithm, Decision Tree Algorithm, Logistic Regression Algorithm, Bayesian Algorithm, Support Vector Machines (SVM), Multi-Layer Perceptron (MLP), and K-Nearest Neighbor (KNN) for training and testing.

When training the models, the training data consists of pre-defined feature matrices and corresponding label matrices. In practice, it is usually necessary to divide the dataset into a training set and a test set, where the training set is used to train the model, and the test set is used to evaluate the trained model. We divided the dataset into a training set and a test set at a ratio of 8:2. Meanwhile, during the training process, grid search is used for hyperparameter tuning, with 5-fold cross-validation. The accuracy rate is selected as the evaluation metric for the model's predictive performance, which is calculated with the following formula:
\begin{equation}Acc=\frac{P_{true}}{P_{all}}\times100\%\end{equation}
where $P_{true}$ represents the number of correctly predicted labels, and  $P_{all}$ represents the total number of predicted labels.

In the process of parameter tuning, the grid search method could improve the generalization ability of the model. The basic idea is to first determine all the network parameters of the machine learning model and their possible candidate values, which are usually given by empirical methods. Then, different parameter grid combinations are generated by enumerating all possible candidate values of the network parameters. For example, if the current model has two hyperparameters a and b with candidate values [a1, a2] and [b1, b2], respectively, the parameter grid combinations would be (a1, b1), (a1, b2), (a2, b1), and (a2, b2). Then, each hyperparameter combination is used to train and evaluate the model, and finally, the hyperparameter combination with the best performance will be selected. The grid search hyperparameter tuning process diagram is shown in Fig. 3.

\begin{figure}[htbp]
\centering
\includegraphics[width=0.7\textwidth]{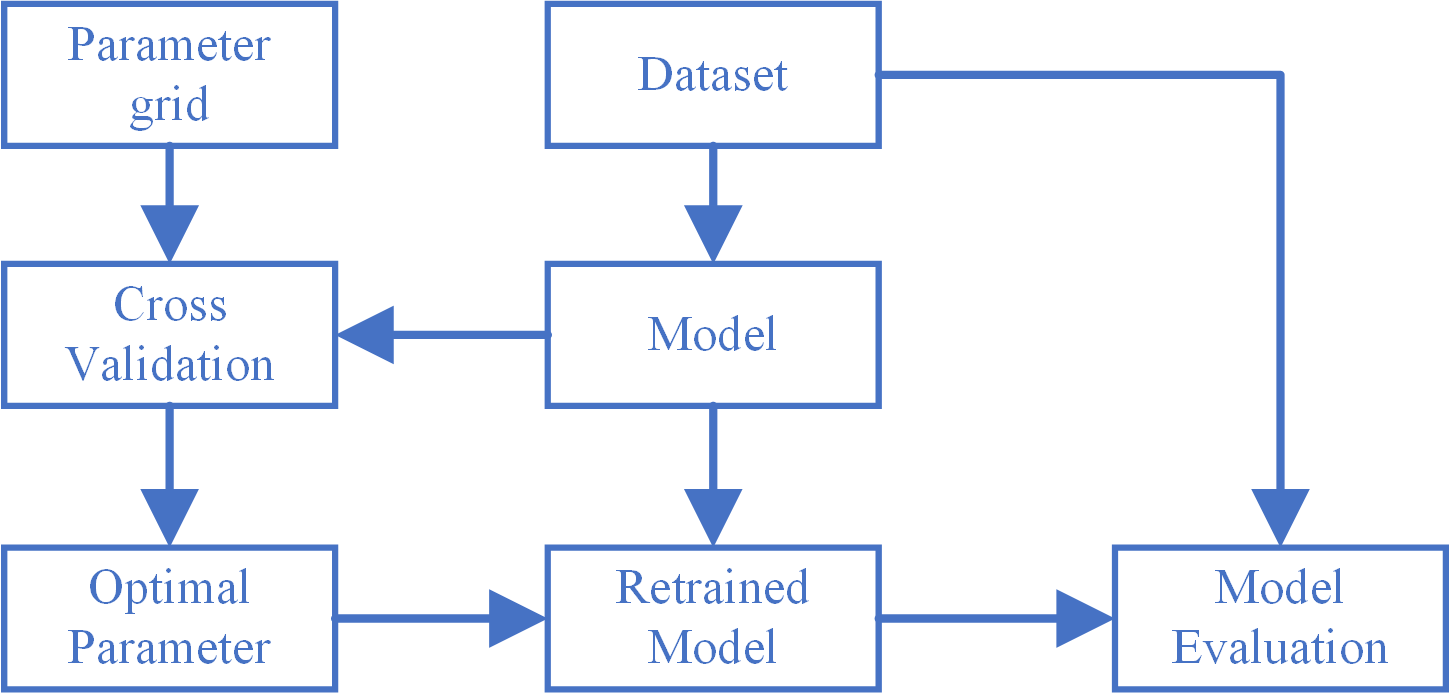}
%\captionsetup\caption
\caption{The parameter tuning process of the grid search method}\label{fig3}
\end{figure}

\section{Experimental Evaluation}\label{sec9}
 \subsection{Experiment Platforms}\label{subsec5}

 This model was trained and evaluated on an AMD Ryzen 5 2600 Six-Core Processor with a clock speed of 3.4 GHz, Python version 3.8.5 and Scikit-learn version 0.23.2, running on a Windows 10 operating system. When solving matrices, the MUMPS version is 5.6.2, compiled with GCC version 11.1.0, and parallelized with OpenMP to utilize 56 threads. This was done on a Linux system with kernel version 5.4.0-65.

  \subsection{Model Prediction Results}\label{subsec6}
  
  To avoid the impact of differences in feature scales on model performance, speed up model convergence, and improve model accuracy, feature normalization is usually performed during model training. There are two common normalization methods: Max-Min Scaling and Standardization. We selected seven machine-learning algorithms from the Scikit-learn library to train and test models. Fig. 4 shows the prediction accuracy of different machine learning algorithms under two normalization methods for sparse matrix reordering algorithm prediction. The Random Forest algorithm achieves the highest accuracy under both normalization methods, making it the best-performing machine learning model. Additionally, most machine learning models achieve higher accuracy when using Standardization for prediction. The experimental results demonstrate that Standardization has a more significant advantage in normalizing sparse matrix features during model training. Therefore, we finally chose Standardization to normalize matrix features and use the Random Forest algorithm for model training and testing. Under Standardization, the prediction accuracy of the Random Forest algorithm reaches 86.7\%, indicating good prediction accuracy. The hyperparameter combination selected by the Random Forest algorithm after grid search optimization is shown in Table 4.

\begin{figure}[htbp]
\centering
\includegraphics[width=0.9\textwidth]{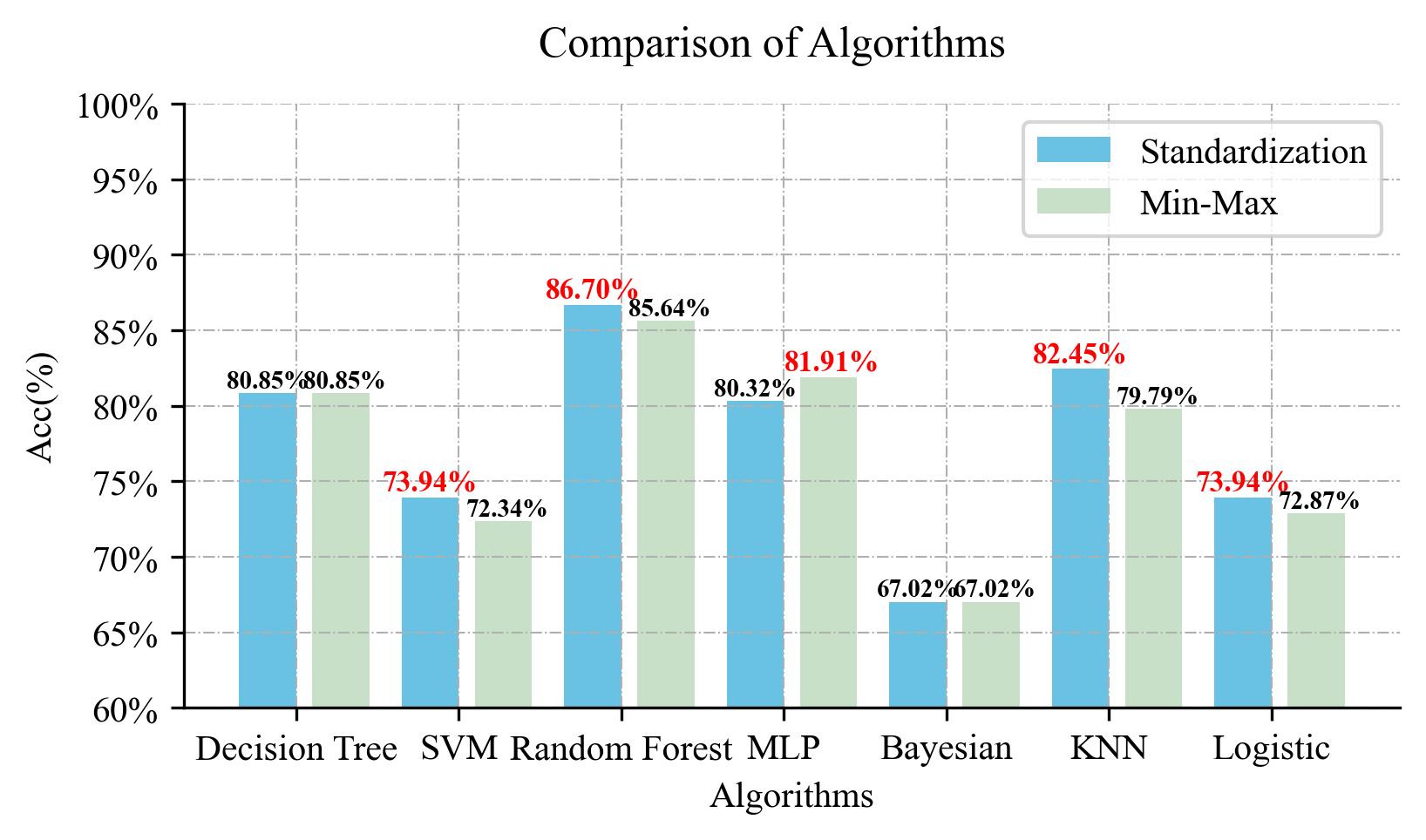}
%\captionsetup\caption
\caption{Prediction Accuracy of Different Machine Learning Algorithms}\label{fig4}
\end{figure}

% \usepackage{booktabs}
\begin{table}[]
\caption{Hyperparameters of the Random Forest Algorithm}\label{tab4}%
\begin{tabular}{@{}cc@{}}
\toprule
Hyperparameter Name & Value \\ \midrule
Criterion           & gini  \\
Min\_samples\_leaf  & 1     \\
Min\_samples\_leaf  & 5     \\
N\_estimators       & 100   \\ \bottomrule
\end{tabular}
\end{table}

To test the model performance for sparse matrix solving problems, we used the trained model to predict the sparse matrix reordering algorithm for the sparse matrices in Table 1 and record the prediction time for each matrix. The prediction results, prediction times, and true labels are shown in Table 5. From the table, it can be clearly seen that the model obtains a very good prediction performance. Moreover, the model's prediction time is almost negligible compared to the time spent solving the matrix, which means that applying the model to the sparse matrix solution process can greatly reduce the time waste caused by using incorrect sparse matrix reordering algorithms, further highlighting the potential practical value of the model in optimizing sparse matrix calculations.
% \usepackage{booktabs}
\begin{table}[]
\caption{Model Prediction Results and Prediction Times}\label{tab5}%
\begin{tabular}{@{}cccc@{}}
\toprule
Matrix Name & Predict Label & Predict Time(s) & True Label \\ \midrule
ASIC\_320k  & ND            & 0.0161       & ND         \\
pf2177      & SCOTCH        & 0.0161       & SCOTCH     \\
crystk02    & SCOTCH        & 0.0161       & SCOTCH     \\
SiH4        & SCOTCH        & 0.0161       & SCOTCH     \\
obstclae    & AMD           & 0.0150       & AMD        \\
lhr07c      & AMD           & 0.0161       & AMD        \\
nemeth17    & RCM           & 0.0151       & RCM        \\
af23560     & AMD           & 0.0160       & AMD        \\
pli         & SCOTCH        & 0.0161       & SCOTCH     \\ \bottomrule
\end{tabular}
\end{table}

We also conducted statistical analysis on the solution time of all matrices in the test set under three scenarios: a) using the traditional AMD algorithm, b) using the machine learning model's predicted reordering algorithm, and c) using the ideal optimal reordering algorithm. The related statistical results of the model's prediction time for all matrices in the test set are shown in Table 6. The solution time results for all matrices in the test set show that, compared to the traditional AMD-based method, the machine learning method can reduce the solution time by 55.37\%. Meanwhile, compared to the ideal situation where all reordering algorithms are optimal, the solution time only increases by 19.86\%. Furthermore, the model's prediction time is almost negligible compared to the solution time. In conclusion, using a machine learning model to predict the reordering algorithm can effectively solve the problem of increased solution time due to poor reordering method selection, while maintaining good performance.

\begin{table}[]
\caption{Statistical Results of Solution and Prediction Times}\label{tab6}%
\begin{tabular}{@{}cccl@{}}
\toprule
\multicolumn{3}{c}{Solution Time(s)} & \multirow{2}{*}{Prediction Time(s)} \\ \cmidrule(lr){1-3}
AMD        & Prediction & Ideal     &                                  \\ \cmidrule(lr){1-4}
2684.3150 & 1198.0040 & 999.5337 & \multicolumn{1}{c}{3.0394}      \\ \bottomrule
\end{tabular}
\end{table}

In addition, the average speedup ratio of the matrices in the test set using the reordering algorithm predicted in this paper is 1.45 compared with the AMD reordering algorithm, and the ten matrices with the largest matrix dimensions in the test set are also selected, and the solution time of the reordering algorithm sorted by the AMD algorithm, the solution time and the speedup ratio predicted by the reordering algorithm in this paper are listed respectively, and the results are shown in Table 7. When the matrix scale is large, the predicted reordering algorithm can significantly accelerate the solution efficiency, and the maximum speedup ratio can reach 25.13, which reflects the advantages of the model in solving large sparse matrices, and can better predict the best reordering algorithm of the matrix, and shows that it is very necessary to find a suitable sparse matrix solution reordering algorithm for large sparse matrix.
% \usepackage{booktabs}
\begin{table}[]
\caption{Performance comparison of the ten largest matrix}\label{tab7}%
\begin{tabular}{@{}cccc@{}}
\toprule
Matrix Name    & AMD(s)      & Model Prediction(s) & Speedup Ratio \\ \midrule
t2em           & 23.4912  & 0.9348           & 25.13         \\
af\_0\_k101    & 105.7766 & 49.6773          & 2.13          \\
NotreDame\_www & 0.1404   & 0.0701           & 2.0           \\
Stanford       & 2.7708   & 0.5386           & 5.14          \\
BenElechi1     & 33.7089  & 26.8109          & 1.26          \\
dc3            & 18.8602  & 1.8920           & 9.97          \\
Torso2         & 1.7802   & 1.7802           & 1.0           \\
Barrier2-9     & 488.8780 & 174.0420         & 2.81          \\
Barrier2-11    & 504.3810 & 184.0130         & 2.74          \\
Barrier2-4     & 432.3810 & 167.2610         & 2.59          \\ \bottomrule
\end{tabular}
\end{table}

\section{Related works}\label{sec2}

In machine learning, the classification task is one of the most widely used tasks in supervised learning, which usually uses statistical learning methods to construct a predictive model that can accurately classify unknown data by learning patterns and patterns in the data based on known training datasets \cite{sharifani2023machine}. The basic idea is to infer and establish a mapping function from the input features to the category labels by training the relationship between the known input features and the corresponding category labels in the dataset. The advantage is that once the algorithm has learned the mapping relationship between labels and data features, the machine learning model can automate decision-making and prediction, and can handle large-scale datasets and complex computational tasks \cite{mahesh2020machine}. These advantages have led to a wide range of applications in various fields.

At present, commonly used and classified machine learning algorithms include decision trees, random forests, multilayer perceptron, and other algorithms. Stylianou et al.\cite{stylianou2023optimizing} proposed a machine learning-based prediction library for sparse matrix format multiplication to select the best sparse matrix format for sparse matrix-vector multiplication. Mehrabi et al. \cite{mehrabi2021learning} improved the performance of SpMM by using a simple two-layer MLP to predict the most appropriate arrangement technique for a given matrix sparse mode. Cui et al. \cite{cui2019machine} proposed a machine learning-based prediction sparse matrix reordering algorithm for matrix-solving problems in the field of power grids, but this is only a simple binary classification model. Tian et al. \cite{tian2023spm} proposed a sparse matrix reordering algorithm classifier based on graph neural networks, but the training is very time-consuming, and the prediction accuracy is only 70\%.

As far as we know, this paper explores the influence of different reordering algorithms on the solution efficiency of sparse matrices for the first time, which provides a new perspective for further optimizing the calculation of sparse matrices.
\section{Conclusions}\label{sec30}
In this paper, the influence of different reordering algorithms on the efficiency of solving sparse matrices is discussed, and it is observed that different sparse matrix algorithms have great differences in the solving performance of the same matrix, and no sparse matrix reordering algorithm can be well adapted to all sparse matrix solutions. Therefore, we proposed a reordering algorithm selection model based on machine learning, which can select a proper reordering algorithm for sparse matrix solving, and predict the most suitable matrix reordering algorithm according to the statistical features of the input matrix, which can realize the intelligence and automation of the matrix reordering algorithm selection and reduce the subjectivity in the process of human intervention and selection. At the same time, the experiments show that the proposed model can accurately predict the optimal reordering algorithm corresponding to different sparse matrix structures from the given sparse matrix reordering algorithm, greatly improving the efficiency of sparse matrix solving.

\section*{Declarations}

\textbf{Conflict of Interest } The authors have no competing interests as defined by Springer, or other interests that might be perceived to influence the results or discussion reported in this paper.

\noindent\textbf{Author contribution } TT generally designed the methods and experiments. JY has conducted experimental testing. The first draft of the manuscript was written by TT and JY, CY and FB has proofread the manuscript. PL and CH and ZP prepared figures All authors read and approved the final manuscript.

\noindent\textbf{Funding } This work was funded by the National Key R and D Program of China 2020YFA0709803.

\bibliography{sn-bibliography}% common bib file
\end{document}